\documentclass[10pt, twocolumn]{article}

\usepackage[utf8]{inputenc}
\usepackage[T1]{fontenc}
\usepackage{lmodern}
\usepackage{booktabs}
\usepackage{tabularx} 
\usepackage{listings}
\usepackage{algorithm}
\usepackage{algpseudocode}
\usepackage{hyperref}

\algrenewcommand\algorithmiccomment[1]{\hfill\textcolor{gray}{// #1}}
\algrenewcommand\algorithmicensure{\textbf{Ensure:}}
\algrenewcommand\algorithmicrequire{\textbf{Require:}}

\usepackage{geometry}
\usepackage{graphicx}
\usepackage{caption}
\usepackage{subcaption}

\usepackage{amsmath, amssymb, amsthm}

\usepackage[numbers]{natbib}
\usepackage{xcolor}
\usepackage{hyperref}
\hypersetup{
    colorlinks=true,
    linkcolor=blue,
    citecolor=blue,
    urlcolor=blue,
}

\usepackage{balance}
\title{Tiny-QMoE}
\author{
    Jack Cashman \\
    \texttt{jcashman@u.rochester.edu}
    \and
    Jiaqi Nie \\
    \texttt{jnie7@u.rochester.edu}
}
\date{}

\begin{document}

\twocolumn[
\maketitle
\begin{abstract}
\noindent
The QMoE model \cite{frantar2023qmoe} provides a practical approach for compression of massive Mixture-of-Experts (MoE) models. QMoE offers a solution geared towards memory limitations that often reach terabyte scales, and it has the advantage of working with high sparsity models which implicitly lend themselves to compression techniques. QMoE also has the advantage of only taking MoE models into account and does not evaluate its use with non mixture of expert systems. Although this prior attempt focuses on the limitations of large servers with the latest NVIDIA hardware which in the case of the H100 and V100 which have 80 GB of HBM (High Bandwidth Memory), what is not being considered is a significantly more constrained environment, such as in the case of mobile devices which may have in the case of the iPhone anywhere from 4 to 8 GB of unified memory which also needs to be shared with the operating system and additional processes. Although edge devices such as phones and laptops are becoming increasingly more computationally powerful, they are still not close to the level of advanced server machines such as NVIDIA. An additional constraint that we must consider is that of latency. The communication time of sending a request to an LLM server and then getting it back is an additional waiting time that can be removed. We may also want to use LLM technology in environments where there is no reliable network connection. As will be discussed later, while the latency of the internet is removed, the latency of the decompression is incurred. In this paper, we present a solution to this highly constrained memory problem. This takes the form of a re-imagined quantization compression schema and execution such that models which would have normally exceeded the memory requirements of say a 2060 with 6 GB of VRAM. Specifically, Tiny-QMoE works on a variety of LLAMA 3.2 models and does so by quantizing the models into 8-bit versions and then taking said models and storing them in a dictionary based compression format.
\end{abstract}
\vspace{1em}
]

\section{Introduction}
\label{sec:introduction}

LLM's or large language models have in the last few years experienced a meteoric rise in popularity. People have used these models for an unbelievably wide array of natural language and reasoning applications. While demand for these models is high, so to is the computational and memory cost of inference. If a user were to try and run a cutting edge large language model locally on device, they will not be able to run it due to this memory limitation. To get around this issue, the conventional knowledge has been that in order to run large language models locally, models with a lower memory threshold are needed. This production of smaller models comes at a noticeable cost in model performance. Significantly shrinking the number of parameters and aggressive quantization often make models so inaccurate that its output will be unusable by users.

Compared to previous solutions Tiny-QMoE comes at the problem of memory constrain of mobile devices in a novel way. The system was required to make significant changes to prior approaches. The main contributions of this work are summarized as follows:
\begin{itemize}
    \item \textbf{Novel Compression Technique:} Developed a memory efficient approach to run quantized large language models (LLMs) locally without significant degradation in model performance or latency. This effort is accomplished through the storage of quantized LLAMA 3.2 models\cite{touvron2023llama} into a dictionary based compression schema.
    \item \textbf{Preservation of Model Accuracy:} Demonstrated the degree to which aggressive quantization can maintain usability and output quality when applied with our method. The issue of accuracy despite quantization must be accomplished through a data dependent approach rather than simply a naive reduction in value representation. The quantization of smaller models also requires a less aggressive degree of quantization for comparable reduction in performance compared to that of significantly larger and denser models. 
    \item \textbf{Enhanced Accessibility of LLMs:} Enabled the use of cutting edge LLMs on resource constrained devices, bridging the gap between high computational demands and device limitations. Large Language Models should be able to be ran on more constrained devices and in turn enable usage on device with the benefit of working without internet connectivity. An on device model would allow for more data privacy as user prompts will not have to be sent to the web. This privacy can minimize data tracking and the fears of possible data breaches. LLM's on mobile devices may also bring the promise of better energy consumption as many cutting edge arm chips may outperform NVIDIA server scale AI systems in terms of performance per watt. This increase in accessibility requires the rejection of hardware specific code and in our current implementation focuses execution of inference on CPU execution.
    \item \textbf{Performance Evaluation:} As we want to ensure that our models are in fact as reduced in size as we would hope for them to be and as performant both in terms of latency and quality of output we must run evaluations. Our evaluation of extensive experiments showcasing the trade off between model size and performance across various reasoning and natural language tasks. We expect there to be an increase in latency and a decrease in quality of output, but not too much. Our focus given these setbacks comes down to the compression which we aim to maximize. If the model size is less than the memory of a system then it would be able to operate within the respective system which would mean that compression can effectively increase the amount of hardware any model can be used on especially given our hardware agnostic approach.
    \item \textbf{Scalable Deployment of Variable Models:} We design our framework with the intentionality that it can be applied to LLAMA 3.2 models of varying size and capabilities. The approach allows for the goals and limitations of the different implementations to be scalable allowing users to opt between the range of models. 
    
    Our range of LLAMA 3.2 models consists of the 1B and 3B models but we plan to expand to include 11B and 90B where each model is named based on the number of parameters they have where LLAMA 3.2 1B has one billion parameters and the rest follow respectively. We have an expectation that as the number of parameters increases the compression rate will also increase.
\end{itemize}

Our proposed current work sadly comes with several limitations however. This in turn makes it the case that simply taking the implementation directly from QMoE\cite{frantar2023qmoe} would not be possible. QMoE simply cannot suit the altered context of these both smaller and non MoE models. While the QMoE approach demonstrates significant advancements in compression and efficiency, these gains face certain limitations associated with our approach:
\begin{itemize}
    \item \textbf{Dependence on High Sparsity:} QMoE relies heavily on high sparsity to achieve substantial compression rates. This dependency makes it less effective for smaller and in turn less sparse models to be used limiting its applicability in certain scenarios like that of LLAMA 3.2 and other smaller to medium sized models.

    The dependency on high sparsity is reasonable in the MoE context. The mixture of experts allows for an increased parameter distribution through a subnet of input tokens. Given that it is the case that smaller and less dense models are less resistant to quantization noise (Frantar et al., 2022) our significantly smaller models cannot be expected to be nearly as resistant to this performance degradation.
    
    \item \textbf{Hardware Specific Implementation:} The compression schema in QMoE is implemented using CUDA, which restricts its use to CUDA compatible devices. This serves as a limitation for hardware agnostic or diverse hardware environments.

    Our goal for implementation is to make system that can run reliably on nearly any CPU architecture. This will come at the performance cost which can be alleviate by future work in the form of a more hardware specific approach that may use a solution such as OpenCL or in the case of Apple hardware OpenXLA.
\end{itemize}

\section{Background}
\label{sec:background}

\subsection{Quantization}
Quantization is a technique which can be used to reduce the precision of numerical representations in machine learning models, typically converting floating point parameters to lower bit integers. This approach reduces memory usage and computational cost while aiming to maintain model accuracy. Quantization plays a critical role in enabling the deployment of large language models on resource constrained devices. Quantization can also help to improve the computational performance of a model as the lower bits enable more parallelization and faster compute.

For the training of models, it makes the most sense to train with high accuracy so that the adjustments of weights can be as precise as possible. High precision ensures that these updates are accurate, which is essential for minimizing the loss function and converging to an optimal solution.

Training involves computing gradients, which can be sensitive to small changes. High precision calculations reduce numerical instability, especially in deep networks, preventing gradients from vanishing or even exploding.

The quantization used in that of QMoE was ternary quantization. The ternary representation consists of using only three possible values. The compression looks at each row of weights and enforces that each weight must be the minimum weight, maximum weight, or zero. Given a near normal distribution this representation inherently causes high sparsity of nearly ninety percent in the case of QMoE.

\subsection{Compression}
Compression techniques aim to reduce the size of machine learning models, improving memory efficiency and computational performance. We achieve this through an LZW-based schema, a dictionary based compression algorithm that dynamically builds a dictionary of substrings encountered in the input data. LZW replaces these substrings with shorter numerical codes, significantly reducing the size of data with repeated patterns.

During encoding, LZW starts with a dictionary containing single character substrings. It reads the input string, identifies the longest substring already in the dictionary, outputs its code, and adds a new substring to the dictionary. When applied to machine learning models, LZW leverages recurring patterns in model weights and parameters to achieve efficient compression. It will be the case that when quantization is more aggressive, so too will the compression as for in the case of ternary quantization frequently occurring patterns are easier to recognize than eight bit int representation. By exploiting structure and redundancy, LZW helps significantly reduce model size while preserving functionality.

\subsection{Inference}
Inference refers to the process of using a trained model to make predictions or generate outputs. Efficient inference is essential for deploying large language models, especially in real time or low latency scenarios. Compression and quantization directly impact inference performance by reducing memory and computational requirements.

In our work, the inference process fundamentally diverges from conventional approaches that rely on fully expanded, high precision weight matrices. Rather than performing inference on a large, static set of high precision parameters, our method encodes frequently occurring patterns of quantized model weights into a compact dictionary based compression schema. During inference, parameters are not decompressed yet and must do so on a layer by layer basis and turn into the quantized model. This reduces both the memory footprint and the computational overhead required to handle the model’s parameters. This is in stark contrast to traditional inference pipelines, which more often than no demand substantial memory bandwidth and or specialized hardware capabilities.

\subsection{Mixture of Experts (MoE)}
\label{sec:moe}

The Mixture of Experts (MoE) architecture is designed to increase a model’s parameter count in turn boosting its modeling capacity while keeping computational costs nearly constant relative to a standard feed forward network. The core idea is to replicate certain model components multiple times creating a set of experts each responsible for processing only a subset of the input tokens. A dedicated router layer assigns tokens to individual experts based on assignment scores, ensuring that only a small portion of the full network is active for any given token. This selective activation leads to increased model capacity without a proportional increase in computational cost, enabling more efficient scaling of large language models.

\subsection{Sparsity}
Sparsity is the frequency of zero to near zero values in a model. Exploiting sparsity enables significant reductions in memory and computation, which is particularly best seen in large scale models. Sparsity is closely tied to entropy and compressibility because when parameters are predominantly zero or clustered around a small set of values, the overall information content of the weight distribution is reduced. The distribution and variance of data plays a key role in determining the compressibility of data. Models exhibiting higher sparsity tend to have fewer distinct patterns within their parameters, effectively lowering the entropy of the parameter distribution and reducing the LZW ability to achieve an efficient dictionary on long patterns. Lower entropy implies that the model’s weights can be encoded using fewer bits on average, thus enhancing compressibility.

When dealing with large MoE based architectures, aggressive quantization can further amplify this sparsity. Quantization of MoE models in ternary representation makes it that they can have nearly a 90 percent sparsity (Frantar et al., 2022) whereas our quantized models have close to zero. This creates a great deal of structured sparsity for ternary QMoE, where entire subsets of parameters become compressible with minimal fidelity loss. As a result, larger MoEs undergoing heavy quantization not only maintain their expansive capacity but also become more amenable to memory and computational optimizations based on sparse representations. In contrast, significantly smaller models cannot be as aggressively quantized without compromising their representational capacity and performance. Because these models have fewer parameters and often rely on finer grained precision to capture subtle patterns in the data, aggressively pushing weights toward zero might degrade accuracy.

\subsection{Rejecting CUDA and Hardware Acceleration}
CUDA is a parallel computing platform that enables efficient implementation of QMoE's decompression stage on NVIDIA GPUs. However, this reliance on specific hardware in the form of a reliance on CUDA restricts the generalizability of the approach to other hardware environments. Developing a custom CUDA based decompression schema can yield significant latency reductions, as it leverages GPU accelerated parallelism to quickly transform compressed sequences back into usable model weights. By distributing the decompression workload across thousands of GPU threads, the process can be executed at scale, thus minimizing stalls and bottlenecks in the inference pipeline.

On the other hand, implementing the decompression routine on the CPU provides a hardware agnostic solution that is not constrained by proprietary platforms or vendor specific toolchains. By avoiding CUDA, the model becomes more portable, running effectively on a wider array of devices, from desktop CPUs and mobile chipsets to specialized accelerators that do not support CUDA. This expanded compatibility eases deployment concerns and future proofs the approach against shifts in hardware ecosystems. While a CPU based solution may not match the raw speedups offered by GPU acceleration, it allows for broader dissemination, ensuring that the compression decompression pipeline is accessible and maintainable across diverse computing environments. It would also likely be the case that the latency incurred by running any inference on CPU would better mask the decompression latency.

\section{Scaling Back Quantization For Realistic Performance}
\label{sec:quantization}

Ternary quantization, which restricts weights to three discrete values (e.g., {$w_{min}$, $0$, $w_{max}$}), introduces a  high degree of information loss, especially in models with relatively small parameter sizes like Llama 3.2-1B. Such aggressive quantization reduces the representational capacity of the model, severely affecting its ability to capture intricate patterns  in data. The resulting loss in granularity can lead to significant degradation in  performance, as seen in our experiment  where the model failed to generate coherent English responses. This suggests that the limited weight resolution in ternary  quantization disrupts the balance between  precision and generalization, making it  unsuitable for smaller models that rely on  higher precision to maintain functionality.

We chose 8-bit quantization after an  extensive series of experiments involving ternary, 2-bit, 4-bit, 6-bit, and 8-bit  naive quantization methods. Our naive  quantization algorithm operates by first  determining the range of weight values  (minimum and maximum) for each layer of the model. Specifically, the algorithm  calculates a scale factor based on the  difference between the maximum and minimum weight values and maps the weights to a  fixed range of discrete levels determined  by the number of quantization bits. For 8-bit quantization, weights are mapped to 256 discrete levels ($2^8=256$) using this scale. The quantization process is  implemented in a Quantizer class, which includes a \textbf{find\_params} function to compute the scale factor and zero point, and a quantize function to map the weights to their quantized values. If the \textbf{maxq} parameter is set to negative (e.g., for ternary quantization), the  algorithm instead applies a simpler threshold logic.

The algorithm iterates through all the parameter weights of the model, and applies quantization as needed, and runs the
scaling and rounding process to replace the
original weights with their quantized
counterparts. This uniform quantization 
approach does not consider the varying 
importance of different weights, which can
result in significant information loss, 
particularly when using lower bit width 
quantization. During our experiments, while
ternary, 2-bit, and 4-bit quantization 
caused a severe loss of model 
functionality, 6-bit and 8-bit quantization
retained the model’s ability to generate
coherent outputs. Among these, 8-bit 
quantization produced responses with 
superior logical coherence and reasoning.

To further optimize model performance, we 
applied GPTQ (Gradient Post-Training 
Quantization)\cite{ref:gptq}, a data dependent 
quantization method. GPTQ addresses the 
limitations of our naive algorithm by 
adapting the quantization process based on 
the gradient and loss landscape of the 
model, ensuring that the most critical 
weights for model performance are 
preserved. By leveraging a small, 
representative calibration dataset (the C4 
dataset in our experiments), GPTQ 
dynamically fine tunes the quantization 
parameters, resulting in a model with much 
lower information loss. We also 
experimented with using GPTQ to perform a 
more aggressive 4-bit quantization. While
GPTQ was able to mitigate some of the 
performance loss, the resulting 4-bit 
quantized model still failed to match the 
performance of the 8-bit quantized model in
terms of both accuracy and coherence. This
highlights the challenges of aggressive 
quantization, as the reduced bit-width 
introduces significant constraints on the 
model’s capacity to represent nuanced 
information. Consequently, we concluded 
that 8-bit quantization strikes the optimal
balance between compression and 
performance, even with advanced 
quantization techniques like GPTQ.

\begin{lstlisting}[language=Python, caption=Naive Quantization]
import torch
from transformers import AutoModelForCausalLM, AutoTokenizer
import torch.nn as nn

tokenizer = AutoTokenizer.from_pretrained("meta-llama/Llama-3.2-1B")
model = AutoModelForCausalLM.from_pretrained("meta-llama/Llama-3.2-1B")

device = "cuda" if torch.cuda.is_available() else "cpu"
model = model.to(device)

class Quantizer(nn.Module):
    def configure(self, bits):
        if bits == 1.5:
            self.maxq = torch.tensor(-1)  
        else:
            self.maxq = torch.tensor(2 ** int(bits) - 1)

    def find_params(self, x):
        dev = x.device
        self.maxq = self.maxq.to(dev)

        xmin = x.min()
        xmax = x.max()

        if self.maxq < 0:  
            self.scale = xmax
            self.zero = xmin
        else:
            self.scale = (xmax - xmin) / self.maxq
            self.zero = torch.round(-xmin / self.scale)

        self.scale = self.scale.unsqueeze(0)
        self.zero = self.zero.unsqueeze(0)

    def quantize(self, x):
        if self.maxq < 0: 
            return (x > self.scale / 2).float() * self.scale + (x < self.zero / 2).float() * self.zero
        q = torch.clamp(torch.round(x / self.scale) + self.zero, 0, self.maxq)
        return self.scale * (q - self.zero)

quantizer = Quantizer()
quantizer.configure(8)

def quantize_model_weights(model):
    for name, param in model.named_parameters():
        if 'weight' in name:
            with torch.no_grad(): 
              quantizer.find_params(param.data)
              param.data = quantizer.quantize(param.data)

quantize_model_weights(model)
\end{lstlisting}

\section{8-bit Quantization Compression}
\label{sec:compression}

Our compression scheme identifies frequently occurring sequences of quantized parameters and assigns them short codewords in a compression table. Sequences known to the table are replaced by their codewords, while unknown sequences are prefixed with a special codeword before being stored raw. This reduces repeated patterns and overall storage.
\noindent\textbf{Compression Code Below:}

\begin{lstlisting}[language=Python, caption=Finding Frequent Sequences]
def find_frequent_sequences(quantized_model, sequence_length=4, top_k=2**16 - 1):
    sequence_counts = Counter()
    for param in quantized_model.parameters():
        weights = param.flatten().detach().cpu().numpy().astype(np.uint8)
        sequences = (
            tuple(weights[i:i + sequence_length])
            for i in range(len(weights) - sequence_length + 1)
        )
        sequence_counts.update(sequences)
    most_frequent = sequence_counts.most_common(top_k)
    compression_table = {seq: idx + 1 for idx, (seq, _) in enumerate(most_frequent)}
    return compression_table
\end{lstlisting}

\begin{lstlisting}[language=Python, caption=Compressing Model Weights]
def compress_model(quantized_model, compression_table, sequence_length=4):
    compressed_files = []
    param_index = 0
    for param in quantized_model.parameters():
        weights = param.flatten().detach().cpu().numpy().astype(np.uint8)
        weights_length = len(weights)
        compressed_param = []
        i = 0
        while i <= weights_length - sequence_length:
            sequence = tuple(weights[i:i + sequence_length])
            if sequence in compression_table:
                compressed_param.append(compression_table[sequence])
                i += sequence_length
            else:
                compressed_param.append(0xFFFF)
                compressed_param.extend(sequence)
                i += sequence_length
        remaining_weights = weights[i:]
        if remaining_weights.size > 0:
            compressed_param.append(0xFFFF)
            compressed_param.extend(remaining_weights)
        compressed_param = np.array(compressed_param, dtype=np.uint16)
        filename = f'compressed_weights_param_{param_index}.npy'
        np.save(filename, compressed_param)
        compressed_files.append(filename)
        param_index += 1
    return compressed_files
\end{lstlisting}

Decompression reverses this process: codewords are mapped back to their original sequences using the table, and raw values are directly restored. This ensures that when reconstructed, the model’s parameters match the original exactly, preserving accuracy while lowering memory requirements.
\noindent\textbf{Decompression Code Below:}

\begin{lstlisting}[language=Python, caption=Decompressing Model Weights]
def decompress_model(compressed_files, compression_table, sequence_length=4):
    decompression_table = {idx: seq for seq, idx in compression_table.items()}
    decompressed_weights = []
    for filename in compressed_files:
        compressed_data = np.load(filename)
        i = 0
        while i < len(compressed_data):
            codeword = compressed_data[i]
            i += 1
            if codeword == 0xFFFF:
                # Read raw values
                raw_values = compressed_data[i:i + sequence_length].astype(np.uint8)
                decompressed_weights.extend(raw_values)
                i += sequence_length
            else:
                sequence = decompression_table[codeword]
                decompressed_weights.extend(sequence)
    return np.array(decompressed_weights, dtype=np.uint8)
\end{lstlisting}

This compression methods which is used on the eight bit quantized llama 3.2 models produces strong results. As can be seen in table 1 when these models undergo the process of quantization and compression their total size becomes a fraction of what it once was. Assuming that the performance of the models both in terms of latency and accuracy remains which in that case would indicate that we may be able to use larger models in more constrained cases especially in the cases of the 11B and 90B models which we plan to test in future work.

\begin{table}[ht!]
\centering
\small
\begin{tabular}{lc}
\toprule
\textbf{Model} & \textbf{Size} \\
\midrule
llama3.2-1B & 2858 MB \\
llama3.2-1B Quantized & 1469 MB \\
llama3.2-1B Quantized+Compressed & 125.29 MB \\
\midrule
llama3.2-3B & 6584 MB \\
llama3.2-3B Quantized & 3522 MB \\
llama3.2-3B Quantized+Compressed & 187.97 MB \\
\bottomrule
\end{tabular}
\caption{Compression Results for 1B and 3B Models with roughly a 23 and 35 times compression rate respectively.}
\label{tab:compression-results}
\end{table}

\section{Experiments}
\label{sec:experiments}

In this section, we benchmark to assess accuracy and latency. Each configuration undergoes 8-bit quantization and an additional compression step. We compare the models:

\begin{itemize}
\item \texttt{llama3.2-1B}
\item \texttt{llama3.2-1B Quantized}
\item \texttt{llama3.2-1B Compressed}
\item \texttt{llama3.2-3B}
\item \texttt{llama3.2-3B Quantized}
\item \texttt{llama3.2-3B Compressed}
\end{itemize}

All latency measurements are recorded on the same hardware configuration, consisting of an Intel Xeon Gold 6130 CPU @ 2.10GHz and a Tesla V100-SXM2 GPU with 32GB of memory, averaging results over a fixed number of samples. The primary metrics we consider are model size, accuracy, and per example latency across three core evaluations: MMLU, ARC-Challenge, and ARC-Easy.

\begin{itemize}
\item \textbf{MMLU (5-shot)\cite{hendrycks2021mmlu}:}Tests the model’s ability to answer multiple choice questions on a broad range of subjects at a college level. The evaluation uses the test set of the MMLU benchmark to assess performance. In the 5 shot setting, a small set of demonstration examples is provided to guide the model in understanding the task format and context. We report both accuracy and latency as key performance metrics.

\item \textbf{ARC-Challenge\cite{clark2018arc}:}
  Tests the model on the MMLU benchmark’s test set, which spans multiple college level subjects. Accuracy is calculated as the percentage of correct answers based on the ground truth. Few shot evaluation is also supported, where a small set of demonstration examples from the validation set guides the model.

\item \textbf{ARC-Easy\cite{clark2018arc} :}  
  A challenging subset of the ARC dataset, focusing on difficult science questions. Correctness is measured by comparing the predicted answers with the provided ground truth. 
\end{itemize}
The system is built using PyTorch and the Hugging Face Transformers library, providing modular support for the three benchmarks:
\begin{enumerate}
    \item \textbf{Model Loading:} Pre-trained models are loaded, with optional replacement of linear layers with compressed counterparts for reduced model size.
    \item \textbf{Dataset Handling:} Each benchmark’s dataset is fetched and prepared, with optional few shot prompts generated using sampled examples from validation or test sets.
    \item \textbf{Evaluation Pipeline:} For each question:
    \begin{itemize}
        \item Prompts are generated and tokenized.
        \item The model computes the log likelihood for each answer option.
        \item The option with the highest score is selected as the prediction.
        \item Accuracy is updated based on correctness, and latency is measured per question.
    \end{itemize}
\end{enumerate}

Latency measurements during inference ensure precise performance metrics, providing a comprehensive evaluation across \textbf{MMLU}, \textbf{ARC-Challenge}, and \textbf{ARC-Easy}. This setup, executed on the specified hardware configuration, ensures no data leakage and robust assessment of the model's performance.

\noindent\textbf{Results:}
The table below presents the model sizes (after quantization and compression), along with the accuracy and latency results on MMLU (5-shot), ARC-Challenge, and ARC-Easy. The quantized and compressed models aim to achieve similar accuracy to the uncompressed baseline while significantly reducing model size and potentially altering latency characteristics.

\begin{table}[ht!]
\centering
\resizebox{\columnwidth}{!}{%
\begin{tabular}{lcc}
\toprule
\textbf{Model} & \textbf{MMLU(5-shot) (\%)} & \textbf{Latency (s)} \\
\midrule
llama3.2-1B & 29.3 & 0.1346 \\
llama3.2-1B Quantized & 29.25 & 0.2113 \\
llama3.2-1B Compressed & 29.25 & 0.2114 \\
\midrule
llama3.2-3B & 35.34 & 0.3292 \\
llama3.2-3B Quantized & 35.31 & 0.5594 \\
llama3.2-3B Compressed & 35.31 & 0.5575 \\
\bottomrule
\end{tabular}%
}
\caption{MMLU (5-shot) results and latency for selected llama3.2 configurations. Accuracy is reported in \%, latency in seconds.}
\label{tab:mmlu-results}
\end{table}

\begin{table}[ht!]
\centering
\resizebox{\columnwidth}{!}{%
\begin{tabular}{lcc}
\toprule
\textbf{Model} & \textbf{ARC-Challenge (\%)} & \textbf{Latency (s)} \\
\midrule
llama3.2-1B & 33.7 & 0.0922 \\
llama3.2-1B Quantized & 33.7 & 0.2609 \\
llama3.2-1B Compressed & 33.62 & 0.2733 \\
\midrule
llama3.2-3B & 57.85 & 0.2504 \\
llama3.2-3B Quantized & 57.59 & 1.3574 \\
llama3.2-3B Compressed & 57 & 1.2866 \\
\bottomrule
\end{tabular}%
}
\caption{ARC-Challenge results and latency for selected llama3.2 configurations. Accuracy is reported in \%, latency in seconds.}
\label{tab:arc-challenge-results}
\end{table}

\begin{table}[ht!]
\centering
\resizebox{\columnwidth}{!}{%
\begin{tabular}{lcc}
\toprule
\textbf{Model} & \textbf{ARC-Easy (\%)} & \textbf{Latency (s)} \\
\midrule
llama3.2-1B & 53.24 & 0.1005 \\
llama3.2-1B Quantized & 52.9 & 0.339 \\
llama3.2-1B Compressed & 52.27 & 0.3191 \\
\midrule
llama3.2-3B & 73.23 & 0.1987 \\
llama3.2-3B Quantized & 72.94 & 1.0164 \\
llama3.2-3B Compressed & 72.56 & 1.1381 \\
\bottomrule
\end{tabular}%
}
\caption{ARC-Easy results and latency for selected llama3.2 configurations. Accuracy is reported in \%, latency in seconds.}
\label{tab:arc-easy-results}
\end{table}

These results provide insight into the trade-offs between compression and performance. Quantization and compression can substantially reduce model size while maintaining competitive accuracy on MMLU, ARC-Challenge, and ARC-Easy tasks, although the impact on latency and final accuracy depends on the model size and the level of compression applied. In particular, comparing the compressed models against their uncompressed and simply quantized counterparts will guide us in understanding how compression interacts with quantization for real world deployment scenarios. It is also important to note that accessing online language models using the likes of \href{https://chatgpt.com/}{ChatGPT} takes hundereds of miliseconds which when I used the developer tools to measure latency on safari for a request it incurred 697ms which is much higher than the latency incurred through the decompression of the on device models.

\section{Conclusion}
\label{sec:conclusion}
The work we presented is Tiny-QMoE, a full scale quantization and compression system designed for LLAMA 3.2 inference in memory constrained systems. We show that models much larger than one would expect to be able to run on these constrained systems can and with minimal performance and latency impact. We show a 23x and 35x compression rate which we assume should continue to be greater for the compression of larger models. We do so in a custom inference execution which decompressed on a per layer basis as to get the most out of the compression. This should make all LLAMA 3.2 models easily more widely available for all as well as at a lower computational cost.

We strongly believe that this work can be furthered. In the future, we will test compression on larger models which was too memory constraining for the system we used. We also hope to test on a larger array of devices as the use of CPU compute would make the system widely extensible into many computing platforms. On top of this while we were unable to bring the model outside of the terminal which we had hoped to do with Web-GPU, we hope to do so in the future as to make this work more public beyond the terminal. The code used for compression, quantization, inference, and evaluation can be found here: \href{https://github.com/Peter-Nie2003/Tiny_QMOE}{Tiny QMoE GitHub Repository}.




\bibliographystyle{plainnat}


\balance

\end{document}